\documentclass{article}

\usepackage{arxiv}

\usepackage[utf8]{inputenc} 
\usepackage[T1]{fontenc}    
\usepackage{hyperref}       
\usepackage{url}            
\usepackage{booktabs}       
\usepackage{amsfonts}       
\usepackage{nicefrac}       
\usepackage{microtype}      
\usepackage{lipsum}		
\usepackage{graphicx}
\usepackage{natbib}
\usepackage{doi}

\title{Utility-Based Context-Aware Multi-Agent Recommendation System \\ for Energy Efficiency in Residential Buildings}

\author{\hspace{1mm}Valentyna Riabchuk  
		\\
	Humboldt-Universität zu Berlin\\
	Berlin, Germany \\
	\texttt{valentyna.riabchuk@hu-berlin.de} \\
	\And
	Leon Hagel \\
	Humboldt-Universität zu Berlin\\
	Berlin, Germany \\
	\texttt{leon.hagel@hu-berlin.de} \\
	\And
	Felix Germaine \\
	Humboldt-Universität zu Berlin\\
	Berlin, Germany \\
	\texttt{felix.germaine@hu-berlin.de} \\
	\And
	\href{https://orcid.org/0000-0003-3506-4744}{\includegraphics[scale=0.06]{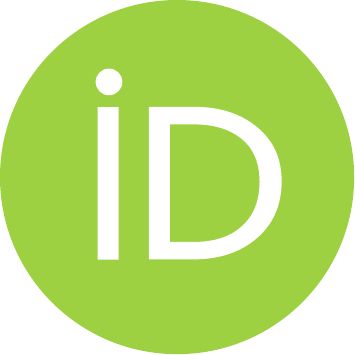}\hspace{1mm}Alona Zharova\thanks{Corresponding author.} } \\
	Humboldt-Universität zu Berlin\\
	Berlin, Germany \\
	\texttt{alona.zharova@hu-berlin.de} \\
}

\renewcommand{\shorttitle}{Utility-Based Context-Aware Multi-Agent Recommendation System}

\hypersetup{
pdftitle={titel},
pdfsubject={subject},
pdfauthor={authors},
pdfkeywords={First keyword, Second keyword, More},
}

\begin{document}
\maketitle

\begin{abstract}
	A significant part of CO\textsubscript{2} emissions is due to high electricity consumption in residential buildings. Using load shifting can help to improve the households' energy efficiency. 
	To nudge changes in energy consumption behavior, simple but powerful architectures are vital. 
	This paper presents a novel algorithm of a recommendation system generating device usage recommendations and suggests a framework for evaluating its performance by analyzing potential energy cost savings. 
    As a utility-based recommender system, it models user preferences depending on habitual device usage patterns, user availability, and  device usage costs. As a context-aware system, it requires an external hourly electricity price signal and  appliance-level energy consumption data. Due to a multi-agent architecture, it provides flexibility and allows for adjustments and further enhancements.
	Empirical results show that the system can provide energy cost savings of 18\% and more for most studied households.
\end{abstract}

\keywords{Recommendation System \and Energy Efficiency \and Load Shifting \and Energy Consumption Behavior}

\section{Introduction}

The energy transition expects the greenhouse gas emissions to be brought to zero by the second half of this century \citep{EU}. To decarbonize the energy sector, the use of renewable energies in the power grid should considerably increase. However, green energy is produced when the weather conditions are met, and therefore is not constantly available in the grid \citep{Dusparic2017}. It should be either stored or consumed when it is available. In addition, the increasing individual electricity demand poses further problems to the grid infrastructure \citep{Giri2009}. 
The obvious solution to tackle these challenges might be to invest in new infrastructures such as storage capacity, additional power plants, or new grid infrastructure. However, this is much more expensive than increasing the energy efficiency through intelligently leveraging the consumption \citep{Palensky2011}. 
The efficient use of energy can contribute to the stability of energy systems through load curtailing and load shifting. While reducing electricity consumption (load curtailing) is not feasible for many private households, shifting the load by reallocating the consumption to better match supply and demand represents a viable approach for real-life applications.
To implement this approach in practice,
the households need data-driven decision support 
from simple but powerful architectures. To this end, recommendation systems provide a valuable framework by recommending energy-efficient actions.


Recommendation systems rely on different algorithms and architectures. Prior work has explored various designs of recommendations systems for energy efficiency in buildings:  content-based \citep{Luo2017}, goal-based context-aware \citep{Sardianos2020, Sardianos2021}, collaborative filtering-based \citep{Luo2021}, fuzzy logic-based \citep{Khalid2019}, user behavior-based \citep{Kar2019, MachorroCano2020}, genetic-based \citep{Yuce2016}, micro-moments-oriented \citep{Varlamis2022, Himeur2022, Alsalemi2019b}, and multi-agent systems \citep{JimenezBravo2019, Pinto2019, Li2015}.
To our best knowledge, there has been limited work on device usage recommendation systems for energy efficiency via load shifting in residential buildings. An extension of traditional recommender systems to device usage recommendations for load shifting is far from straightforward and requires alternative approaches. The  task's difficulty  mainly resides in the complexity of the item to be recommended (i.e., a device usage schedule) and the high dependence of user preferences on context (e.g., user presence, user availability, varying energy prices). These might be the reasons the topic has only been sparsely explored.



This paper proposes a novel recommendation algorithm for generating device usage recommendations for load shifting. As a utility-based recommender system, it models user preferences depending on habitual device usage patterns, user availability, and  device usage costs. As a context-aware system, it requires an external hourly electricity price signal for the following day and historical household electricity consumption data at the appliances level. Based on this data, the recommendation system suggests devices' starting times for the next day with hourly precision that would reduce usage costs while taking into account user habits and availability. 
Due to the multi-agent architecture, 
our system provides additional flexibility and easily allows adjustments and further enhancements.
In addition, we suggest a framework for evaluating the performance by analyzing potential energy cost savings. Besides measuring the system's performance in total, we show how to determine the individual agents’ performance, quantify the cold start problem of the recommender system and analyze the system’s sensitivity to changes in the recommendation hyperparameters. 

We provide a statistically grounded solution for predicting user availability to make recommendations when the user is at home and available for recommendations. The approach also uses a probability of the device usage on a given day to decide whether to recommend or not.  As a result, our recommender system flexibly adapts to actual usage patterns or changing usage behavior. Moreover, it can provide multiple recommendations per week for frequently used devices and few (or none) for infrequently used ones. In addition, the provided algorithm allows calculating costs with higher precision since the system takes into account devices’ load profiles to compute an overall estimation of the usage costs. In other words, it considers not only the electricity price at the device's starting hour but also the costs for the following usage hours, given the multihour device's usage duration.

The empirical analysis aims to quantify how much a user can save by utilizing our recommender system. To achieve this, we create recommendations, determine the acceptability of the recommendations and calculate the energy costs for running the devices with and without recommendations to receive the potential energy cost saving. The results show that the system can provide energy cost savings of 18\% and more for most studied households if the user implements all recommendations.

We summarize the contributions of this paper as follows. 
Firstly, we propose a novel utility-based context-aware multi-agent recommendation system for energy efficiency via load shifting in residential buildings. Secondly, we suggest a framework for the
performance evaluation of the device usage recommendation system. Thirdly, to our knowledge, we are the first to provide a formalization of the task for the device usage recommendations for load shifting.

We provide a comprehensive tutorial in Jupyter Notebook with code in Python for all the steps described in this paper and beyond. The Jupyter Notebook is available on GitHub [link to be added after the review process]. 


The paper is structured as follows. Section \ref{RelatedLiterature} provides an overview of related literature. Section \ref{RecSys} introduces our recommender system approach and the framework for its performance evaluation. We apply the system to the real-world data, analyze its performance and discuss the results in section \ref{EmpiricalResults}. Section \ref{Discussion} elaborates on strengths and weaknesses and discusses future work, whereas section \ref{Conclusions} concludes this paper.


\subsection{General Framework for Device Usage Recommendations for Load Shifting}
\label{section:formalization}

We start by proposing an
abstract characterization 
of the task of making device usage recommendations for load shifting. 
We describe a recommendation as a pair consisting of a recommendation time (i.e., when the recommendation is displayed to the user) and a recommendation item (i.e., what is being recommended to the user). We see a recommendation item as a  suggestion of a usage schedule for a given number of devices over a given time horizon, as described in Table \ref{tab1}.

\begin{table}[h!]
    \centering
    \begin{tabular}{ l l  }
        \toprule
     Device & Recommended time window for the device launch \\
     \midrule
      Washing-Machine  & 11.05.2021 10:12:01 to 11.05.2021 11:12:01 \\
      Tumble-Dryer  &  11.05.2021 11:22:31 to 11.05.2021  12:12:36 \\
     \toprule
    \end{tabular}
    \vspace{0.5cm}
    \caption{Example of a recommendation item of a device usage recommender system.}
    \label{tab1}
\end{table}

An ideal recommender system would recommend such schedules for long time horizons, taking all household devices into account and providing precise usage time windows while matching users' preferences perfectly and hence considering contextual information. Beyond suggesting optimal usage times, this might also mean providing the recommendations at the right time and avoiding suggestions that the users would execute on their own.

Obviously, achieving this ideal is unrealistic in practice. However, we can consider practical implementations as special cases of this ideal that (i) put restrictions on the complexity of the item to be recommended, (ii) put restrictions on the timing of the recommendation, and (iii) simplify the modeling of users' preferences as well as related contextual information. For instance, restrictions might be imposed on the type of devices (e.g., only the washing machine), the recommendation time horizon (e.g., recommend a schedule for the next ten minutes or the next day), the precision of the time window for a device usage (e.g., specify the exact minutes, or recommend use within a given hour or day). Furthermore, the point in time when the recommendation is provided might be fixed to a certain period of the day or be conditional on a particular event (e.g., a user requesting a recommendation via a smartphone). Finally, the recommender system might only aim to capture restricted aspects of user preferences and contextual information (e.g., only consider preferences regarding energy prices and user availability).

While these aspects describe device usage recommendations in general, they do not explicitly describe recommender systems specifically in the context of load shifting for energy efficiency. Given the general framework, we interpret the additional nuance as considering contextual electricity demand and supply information in the modeling of users' preferences. For example, the recommender system could include electricity prices or renewable supplies in the users' preference evaluation. 

\section{Related Literature}
\label{RelatedLiterature}
\subsection{Recommendation System Architectures for Energy Efficiency}

Energy efficiency term implies that consumers use electricity efficiently and, as a result, contribute to the stability of energy systems. Two definitions clarify efficient energy consumption: load curtailing and load shifting. Load curtailing means a reduction of energy use \citep{Finn2013}. For instance, \citet{Behl2016} apply load curtailing in large buildings to reduce the overall buildings' energy consumption in return for a financial reward. Load shifting, in turn, suggests utilizing the same amount of energy but reallocating the consumption to better match supply and demand \citep{Bartusch2014, Tuomela2021}.  
Such reallocation means increasing electricity consumption during periods with high green energy supply or low load in the electricity grid and at the same time reducing energy use during other periods. 
Periods with more green energy and low load in the power grid relate to low market prices for electricity. 
Using price-based or other signals nudges consumers to change their energy consumption behavior \citep{Liu2019, He2021}.
For example, dynamic day-ahead tariffs encourage consumers to use appliances when the energy price is low and, thus, flatten the demand peak on the energy market and relieve the power grid  \citep{Klaassen2016, Dusparic2017, Vanthournout2015}.


Recommendation systems can help users in controlling and changing their electricity consumption behavior. 
Such systems rely on various algorithms and architectures to generate recommendations for energy management. While some recommender systems aim at load curtailing, others target load shifting by using the appliances at the suggested time slots. However, these two types of recommender systems use almost the same data (appliance-level electricity consumption, occupancy data, etc.). Another categorization recognizes recommender systems for private households and commercial (or large) buildings. 

Recommendation systems of various designs aim to reduce energy consumption in private households. 
\citet{Sardianos2020} propose an architecture to the goal-based context-aware recommender system that supports users' energy-saving habits. Their system utilizes energy consumption, occupancy, and outdoor temperature data to create recommendations (i.e., switch off the heating on exit or switch off heating when warm enough).
\citet{Luo2017} create an energy-saving appliance recommender system based on a content-based recommendation technique. This system recommends  ads for energy-saving appliances to the users. First, the approach creates user-based utilization profiles on the appliance level with the non-intrusive load monitoring (NILM) technique to find the users’ needs and interests on the household appliances. Second, after calculating the similarity of the user profile and the item (appliance ad) profile, the system shows the n-top ads to the user. Thus, the authors claim it contributes to sustainable demand and potentially create opportunities to save energy.
An energy management controller based on the fuzzy logic by \citet{Khalid2019} targets the reduction of energy consumption, costs, and peak-to-average ratio. This system uses occupancy data in the household, temperature, and electricity price to control lights, heating ventilation and air conditioning and schedule the appliances' usage. 
\citet{Kar2019} develop a recommender system for intelligent building lighting control that learns user preferences, luminance sensor data, and energy consumption from historical data. The system offers recommendations about lighting intensity for the occupant’s position in the building on the smartphone interface. 
\citet{MachorroCano2020} present an IoT home management system for home comfort, safety, and energy-saving. The J48 machine learning algorithm and Weka API analyze and learn user behavior and energy consumption patterns. The energy-saving recommendations for households can also call some services if necessary (e.g., police, electrical repairs).

Load curtailing recommender systems for commercial or large buildings mainly develop more general recommendations on electricity consumption than systems for individual consumers. 
\citet{Pinto2019} introduce a case-based multi-agent recommender system for large buildings for energy-saving consumption. This system recommends the amount of energy reduction compared with past savings. The k-nearest-neighbor clustering algorithm uses past data to find cases similar to the target. Then, a support vector machine (SVM) technique sets optimal weights for case parameters. The case parameters consist of information on the electricity consumption in building and electricity generation, the number of people, external temperature, price tariffs, etc.
\citet{Wei2020} develop a recommender system for large commercial buildings to change location from one room to another or shift the arrival or departure schedule in the building to a set amount of time earlier or later. The system utilizes data on energy-consuming resources (heating, ventilation, air conditioning, lighting, electric loads), spaces, and occupants. It uses basic descriptive statistics and customized algorithms to implement the approach.

Only a few recommender systems or device usage scheduling architectures aim at load shifting in private households. 
An artificial neural network (ANN) with a genetic algorithm by \citet{Yuce2016} creates an optimal weekly appliance scheduling in the household to reduce the residential demand and grid energy usage during peak hours and maximize the use of renewable energy produced by the eco-friendly house. This approach utilizes the information on renewable energy generation potential of the house, occupants' behavior (i.e., interactions with their appliances), outdoor air temperature data, and other contextual information. 
\citet{Sardianos2021} and \citet{Varlamis2022} introduce the mechanism of persuasive context-aware recommendations for the efficient consumption. The recommender system proposes the efficient timing for appliance usage based on a presence of users, electronic appliance usage habits, and a context (e.g., indoor and outdoor temperature). 
\citet{Luo2021} propose a personalized recommender system based on a collaborative filtering technique. They simulate the data for 400 households to run the system. First, the system  clusters users into highly responsive and less responsive based on the mean monthly energy consumption and energy costs. Second, it creates users’ lifestyle profiles represented by the usage profiles of non-shiftable appliances. Third, it calculates the similarity of all users. Finally, the shiftable appliances' usage habits from the highly responsive customers are recommended to the less responsive customers. 

Applying multi-agent architecture design to the recommender system can provide additional flexibility through the agents' structure. 
A multi-agent recommender system by \citet{Li2015} consists of the supply-side management (e.g., solar panel, main grid management) and demand-side management (i.e., efficient appliances’ use in the household). The agents work together to reduce energy consumption during the peak period and reduce the consumers’ bills. 
\citet{JimenezBravo2019} develop a multi-agent recommender system that suggests recommendations for shifting appliances  when the appliances' use is most efficient. The recommender system consists of the behavior, crawler, and recommender agents. The behavior agent analyzes each appliance's historical energy consumption data  in a household and forwards the result to the recommendation agent. The crawler agent collects the real-time energy price data and provides them to the recommendation agent as well. The recommendation agent receives the data from the other agents and creates a recommendation for the user. This architecture enables the system to perform multiple tasks in a parallel mode.


\subsection{Challenges for Traditional Recommender Systems}

This section analyzes the existing recommender system types for the task of device usage recommendation for load shifting. 
We aim to shed light on the challenges posed by this task when using traditional recommender systems and to provide an overview of possible approaches for device usage-related recommendations.

\paragraph{Content-Based Filtering}
Content-based filtering  recommender systems make recommendations by matching a user to a set of items based on a description of the item to be recommended (i.e., its content) and a user-specific profile that describes the user's preferences independently of other users \citep{PazzaniBillsus2007, Wang2018}. For instance, a content-based  recommender system could suggest books (i.e., the item) considering author and genre (i.e., the considered content) based on a user's previous book purchases (i.e., the user profile). Similarly, but less intuitively,  a recommender system could suggest a device usage schedule (i.e., the item) considering features such as times of use (i.e., the considered content) based on a user's previous device usage behavior (i.e., the user profile). However, for multiple reasons, the latter application might miss the desired target. On the one hand, the more complex we allow the item or schedule to be (i.e., the precision of usage time recommendation, number of considered devices, time horizon), the more difficult it can be to identify items and schedules the user likes.  On the other hand, the simpler the form of the recommendation item, the less useful the recommendation might be. For instance, if we consider a simple task to recommend a single day of use for a single device within one week, it should be relatively straightforward to identify a usage day that fits the users' preferences. However, it is questionable whether such a recommendation would be useful. In contrast, if all household devices are considered, for a long time horizon, with high usage-time granularity (e.g., specify with minute precision), it would be difficult to identify which type of schedule the user prefers. Indeed, the items or  
schedules would be so diverse that past usage data might not suffice to identify user preferences. In other words, an unrealistically large amount of the user's historical data might be required to make sensible recommendations.  
In short, there is a substantial trade-off between the complexity of the recommendation item and the identifiability of users' preferences (for a given amount of usage data) \citep{PazzaniBillsus2007} that likely renders content-based recommendation systems unfit for our task. 

\paragraph{Collaborative Filtering}
In cases where the recommended item displays high complexity, the collaborative filtering approach often enables leveraging other users' opinions to provide sensible recommendations. This approach utilizes the preferences of other users regarding the same or similar items to provide recommendations \citep{Zhao2021}. Collaborative filtering, however, requires a comparison of the recommended items of different users \citep{Koren2015}, which is difficult in device usage recommendations. For instance, devices of the same type (e.g., washing machine) can have diverse functionalities and electricity load profiles in two different households. Moreover, most items and schedules  are unique in a recommendation system without severe restrictions. This makes the standard collaborative filtering approach that employs a user-item rating matrix useless, as it requires multiple user evaluations per item \citep{Schafer2007}. An alternative might be to cluster unique items into larger groups of similar items. However, this task becomes highly non-trivial when items or schedules are complex. Last but not least, the collection of device usage data  and their sharing between households is challenging due to data privacy reasons.

\paragraph{Knowledge-Based Recommender systems}
Knowledge-based recommender systems involve users in specifying their preferences explicitly \citep{Aggraval2016}. For instance, users may specify constraints on the item to be recommended. In the context of device usage recommender systems, users could specify available  time slots for device usage or energy price thresholds for the usage of particular devices. These user specifications can then be combined with domain knowledge (e.g., rules, constraints, or utility functions) to provide relevant .  
Some examples of such constraints in device usage recommender systems are making recommendations for the user-specified time slots and for user-specified devices only. The recommendation system could also model a user's utility function considering energy prices and discomfort. Such algorithms are especially appropriate when the recommended item has complex varying properties, and the association of sufficient ratings with a large number of combinations is difficult \citep{Burke2000}. 
Therefore, knowledge-based recommender systems have the potential to precisely tackle the issues mentioned above. The challenge with this approach is to adequately use domain knowledge for providing satisfactory recommendations. 

\paragraph{Utility-Based Recommender Systems}
Utility-based recommender systems are closely related to knowledge-based recommender systems, as they use domain knowledge to define an appropriate utility function for providing satisfactory recommendations \citep{Aggraval2016}. Such recommender systems can take many different forms and highly depend on the application at hand. The challenge resides in designing an appropriate utility function \citep{Zihayat2019}, which is  highly domain-specific. An advantage of such a solution is that the integrated domain knowledge can partly compensate for the lack of user data and, therefore, deal with the already mentioned issues of item's complexity and variability. In the context of device usage recommender systems, the existing general aspects of user preferences (i.e., the desire to use cheap or renewable energy or not to change the device usage habits too much) can facilitate the generation of relevant recommendations.  

\paragraph{Context-Aware Recommender systems}
Context is essential for making recommendations for load shifting. According to our characterization of the task, it is necessary to take electricity demand and supply context information into account, by definition. Moreover, user preferences for device usage should heavily depend on contextual information of multiple nature (e.g., presence at home or current availability). 
\cite{Adomavicius2011} suggest the following approaches to consider the context: contextual pre-filtering, contextual post-filtering, and contextual modeling. Contextual pre-filtering and post-filtering can enhance context-unaware recommender systems with contextual filtering \citep{Haruna2017}. For instance, to recommend an item within a given context, the pre-filtering approach uses only item ratings collected in the same or similar context, whereas the  post-filtering approach filters out the output of a context-unaware recommender system based on the context information. In contrast, contextual modeling directly uses the context information in the evaluation of users' item preferences \citep{Raza2019}. For example, in utility-based recommender systems, this means including context information (e.g., electricity price) into the selected utility function (e.g., saving costs).

An extension of traditional recommender systems and existing architectures to device usage recommendations for load shifting in private households is far from straightforward and requires alternative approaches. The difficulty of the task mainly resides in the complexity of the item to be recommended (e.g., a device usage schedule) and the high dependence of user-preferences on context (e.g., user presence, user availability, varying energy prices). These might be the reasons the topic has only been sparsely explored.



\section{Utility-based Context-Aware Multi-Agent Recommendation System}
\label{RecSys}

The goal of the suggested approach is to provide the user with device usage recommendations for energy efficiency and, in particular, load shifting. A recommender system that follows this objective can take many different forms that depend on the recommended time-frame, the granularity of the recommendation times, the number of devices, and the considered context information. 
We propose a  device usage recommender system for load shifting that emerged from the pragmatic objective to provide a recommender system algorithm that can be feasibly implemented in private households and trained on publicly available household data. This means that we have restrictions on (i) the type and amount of data that can be used (no simulations), (ii) the technical equipment necessary for implementation, and (iii) the required user input. However, these constraints enable us to provide a readily implementable solution whose concept can be evaluated in a realistic context.

\subsection{Methodology}
This section  provides a formal characterization of our recommender system. First, we describe the recommendation form by building upon the general framework of the device usage recommender system provided in \ref{section:formalization}. Next, we describe the recommendation algorithm that generates the recommendation.

\subsubsection{Recommendation Form}
Our device usage recommendation consists of a recommendation  item (i.e., a suggested device usage schedule) and a recommendation  time (i.e., when the recommendation is provided). The moment of a new recommendation's generation is fixed at a specific time point for each day. In other words, a device usage schedule for the next day $d$ (i.e., for the next 24 hours) is suggested to the users at fixed time on the current day $d - 1$. The day $d$ is also called prediction or recommendation day. 
The users specify in advance the shiftable devices $S^i$, $i = 1, ..., n$, for which recommendations are made  and the devices that show their availability $A^j$, $j = 1, ..., m$, i.e., devices that require user attention or action. 
The recommender system produces one recommendation per device and generates the device's launching hour $h_{S^i}$ with hourly precision while leaving the stopping time open-ended. The recommendation has the form presented in Table \ref{tab2}.


\begin{table}[h!]
    \centering
    \begin{tabular}{ l l  }
        \toprule
     Shiftable Device ($S^i$) & Suggested Starting Hour ($h_{S^i}$) \\
     \midrule
      $S^1$  &  $h_{S^1}$  \\
      $S^3$  &  $h_{S^3}$  \\
       ...  & ...    \\
      $S^n$  &  $h_{S^n}$ \\
      \toprule
    \end{tabular}
    \vspace{0.5cm}
    \caption{Device usage schedule for shiftable devices $S^i$, $i = 1, ..., n$, for the day $d$, provided at a fixed time on $d-1$.}
    \label{tab2}
\end{table}

\subsubsection{Recommendation Algorithm}

\begin{figure}[h!]
	\centering
		\includegraphics[width=0.6\textwidth]{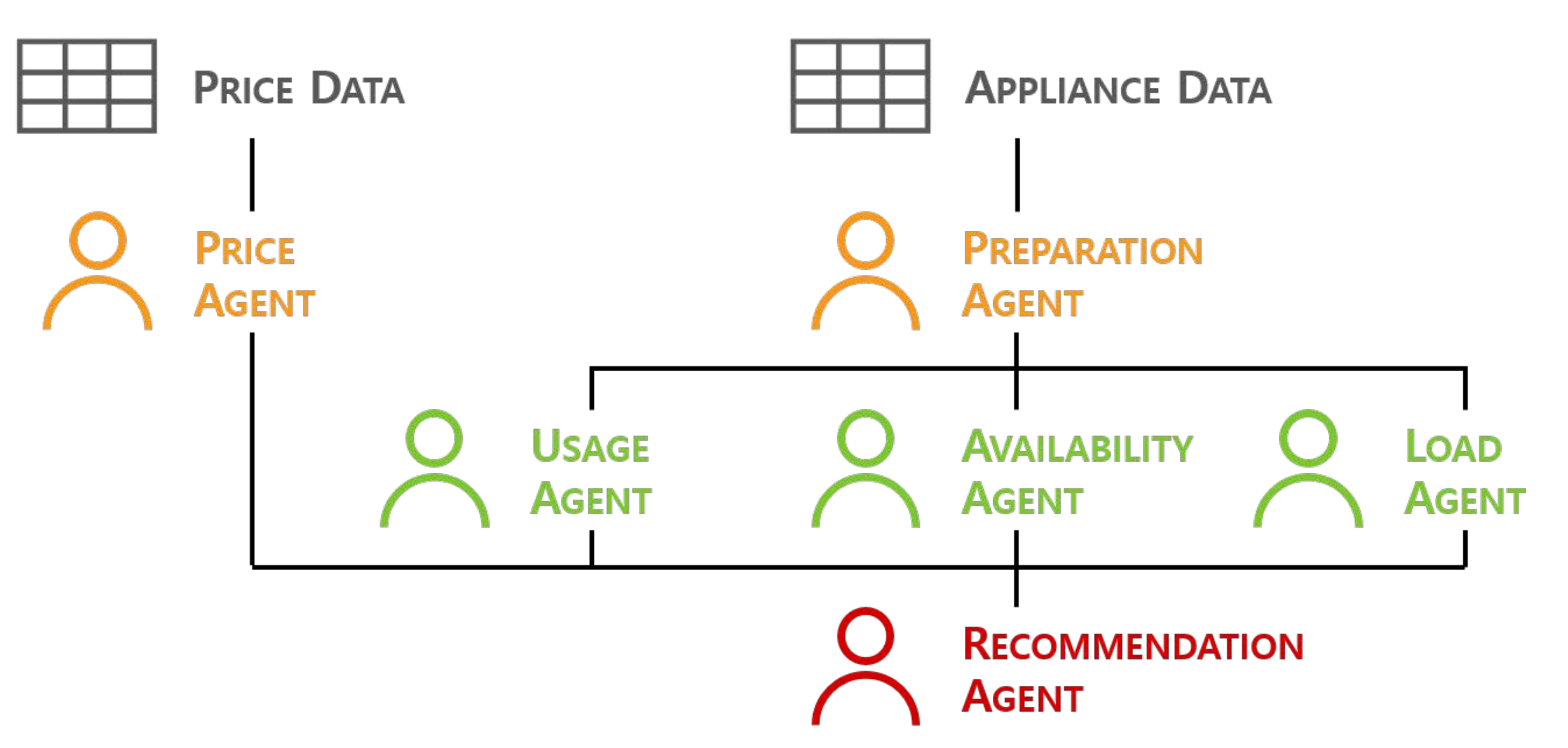}
	\caption{Architecture of the multi-agent approach.}
	\label{fig:multi_agent_1}
\end{figure}

Our algorithm uses a multi-agent architecture (see Figure \ref{fig:multi_agent_1}) to tackle different tasks. 
In a nutshell, the Price Agent provides electricity prices for the following day with hourly precision. The Preparation Agent prepares data for other Agents.
The Recommendation Agent combines the outputs of the Usage Agent, the Availability Agent, the Load Agent and the Price Agent to generate recommendations.

The \textbf{Price Agent} requests price signals and makes them available to the other agents. In particular, it provides a vector of electricity price signals for every hour $h_d \in \left\{0,...,23\right\}$ of the recommendation day $d$ and additional $k$ hours, in the form $p_d = (p_{0}, ..., p_{{23}+k})_d$.
The $k$ hours denote the amount of hours needed for a device to finish its operation in addition to a starting hour. For example, for a washing machine that is running for three hours the $k$ equals to two.
Other hourly day-ahead signals than price can be used here. For the sake of simplicity, we present the system with a concrete example of day-ahead electricity prices.

The \textbf{Preparation Agent} transforms the raw electricity consumption signals to input features for the other agents. A critical task involves detecting when a given device is actually in use, as opposed to noise coming from automated operations. The recommendation system needs to retrieve this information to receive the user availability feature (i.e., availability device used within a given hour), the device usage feature (i.e., specific device used within a day), and typical load profiles. We solve the detection of the device usage periods by setting electricity consumption thresholds above the value that we assume to indicate a device in use. The Preparation Agent requires device-specific electricity load monitoring at a maximum resolution of an hour to provide the required output. The agent monitors both availability devices and shiftable devices. 

The \textbf{Load Agent} provides a  typical load profile $\bar{l}^{S^i} = (\bar{l}^{S^i}_0, ..., \bar{l}^{S^i}_k)$ for every shiftable device $S^i$ based on load data up to that day. 
The typical load profile is computed as hourly averages of all previous hourly load profiles for a device $S^i$ (from device start, with a time window of $k+1$ hours). 

The \textbf{Availability Agent} defines a user as available within an hour of the recommendation day whenever an availability device $A^j$ is used.
It produces user availability probabilities for every hour $h_d \in \left\{0,...,23\right\}$ of the recommendation day $d$, in the form $\pi^U_{d} = (\pi^U_0,..., \pi^U_{23} )_d$. We use a logistic regression model trained on all data up to the day of recommendation to calculate the probabilities. Its predictors consist of time features and user availability lags.

The \textbf{Usage Agent} provides usage probability ${\pi}^{S{^i}}_d = (\pi^{S^1}_d,..., \pi^{S^n}_d) $ for every shiftable device $ S^i$ on the recommendation day $d$. A logistic regression model trained on all data up to that day calculates the probabilities. Its predictors include device usage and user availability on previous days.

The \textbf{Recommendation Agent} provides a recommendation for the next day $d$ for every shiftable device  $ S^i$  from a set of devices $(S^1,..., S^n )$ that have the usage probability greater than a given threshold, $ S = \left\{S^i \mid \pi^{S^i}_d > t_S \right\}$. For such devices, it recommends the cheapest launching hour ($h^{S^i}_d$) among the hours with a probability of user availability greater than a given threshold, $H = \left\{h_d \in \left\{0,...,23\right\} \mid \pi^U_{d} > t_U \right\}$. For each launching hour $h_d$ within the latter set, the system computes the usage costs as a dot product of the typical load profile and the day-ahead prices. Therefore, $h^{S^i}_d = {h_d \in H}{{argmin}} \left((p_{0}, ..., p_{{23}+k})_d \cdot (\bar{l}^{S^i}_0, ..., \bar{l}^{S^i}_k)' \right)$. The system suggests no recommendation to the user, if the probability of device usage or user availability in the recommendation hour is lower than a threshold
. The thresholds  $t_S$ and $t_U$ are initialized at a given value and can be adjusted by the user.


\subsection{Performance Evaluation}
\label{PerformanceTheory}

We suggest a framework for evaluating our recommender system’s performance by analyzing potential energy cost savings. Besides measuring the system's performance in total, we show how to determine the individual agents’ performance, quantify the cold start problem of our recommender system and analyze the system’s sensitivity to changes in the recommendation hyperparameters.

\subsubsection{Evaluating the Performance of Individual Agents}

The agents for which we analyze the individual performance include the Availability Agent, the Usage Agent and the Load Agent. Since the Availability Agent and the Usage Agent perform a standard binary classification task, we use the Area Under the ROC-Curve (AUC score) to evaluate their performance. For calculating the AUC score, we run the pipeline iteratively for every available date $d$ in the data set of the respective household. Hence, we receive 
predictions for user availability and device usage. For each of these predictions, we perform a train-test split and train a new model. To receive a single AUC score, we combine the 
predictions for each agent and compare the predictions to our user availability and device usage targets. 

For evaluating the less standard task of determining the typical load profiles, we use a mean squared error (MSE score). We calculate MSE scores by taking the mean squared error of the actual energy consumption of using a device $(l^{S^i})_{d,j}$ (i.e., true load) in comparison to the current typical load profile $(\bar{l}^{S^i})_d$, where $d$ represents different dates, and $j$ represents different uses within a single day. After calculating the MSE scores for the individual device runs, the scores are aggregated by averaging to receive the MSE score. We use both, the true load on days of use $d$, $(l^{S^i})_{d,j}$ and the typical load profile $(\bar{l}^{S^i})_d$ in their vector representation. Let $n$  be the total number of uses of the device, $n = \sum_d \sum_j1$, and $k$ the usage duration in hours, then the MSE is defined as: 

$$ MSE^{S^i} = \frac{1}{n}\sum_{d}\sum_{j}\frac{1}{k}{((l^{S^i})_{d,j}- (\bar{l}^{S^i})_d )}^T{((l^{S^i})_{d,j}- (\bar{l}^{S^i})_d )}, $$

where $ (\bar{l}^{S^i})_d  $ is a typical load profile of device $ S^i$ at usage date $d$, and $(l^{S^i})_{d,j} $ is a true load profile of device $ S^i$ at usage date $d$ with index $j$. Thus, the MSE score measures how appropriate are the typical load profiles to approximate the true energy consumption of running the respective device.

\subsubsection{Evaluating the Cold Start Problem}

One main challenge related to recommender systems is the cold start problem. It describes a situation when a recommender system might not provide appropriate recommendations due to limited training data at the system's start \citep{Aggraval2016}. Possible solutions to this problem include asking for more user input at the start or transferring knowledge from existing users to new ones \citep{Masthoff2015}. Due to the limited user interaction and the lack of large-scale data, these two approaches are not applicable in our case.


To quantify the cold start problem, we evaluate each agent for varying lengths of its training data and evaluate how the performance improves for larger training data sets. For tracking the performance variations, we need comparability of the test data. Therefore, we evaluate different versions of an agent (one version for each training data length) on the same test data. For the Availability Agent and the Usage Agent, we specify the test data in this way and continue to use the AUC score to evaluate their performance. 

To measure the cold start performance of the Load Agent, more fundamental changes are required. Since the Load Agent does not perform a standard machine learning task, we create a new measure to capture the cold start performance. The goal of the Load Agent is to extract a typical load profile for each shiftable device. Our best estimate for the typical load profile $\bar{l}^{S^i*}$ of a device is the average load profile calculated on the complete data set. This estimate can be approximated by the typical load profiles created by the Load Agent for the last date to be predicted. To analyze the performance over time, we compare the current typical load profile provided by the Load Agent at each date to be predicted to this best estimate. Hence, our performance measure for the Load Agent has to be a distance measure. It has to measure the distance between the typical load profile of the date $(\bar{l}^{S^i})_{d}$ to be predicted and our best estimator of the typical load profile. We can consider the load profiles as a vector representation of the average hourly energy consumption of the shiftable devices. Therefore, we use a normalized Euclidean distance as our distance measure, see:

$$ d_{norm}(\bar{l}^{S^i}_{d}, \bar{l}^{S^i*}) =  \frac{\left\| (\bar{l}^{S^i})_{d}-\bar{l}^{S^i*} \right\|_2}{\left\|\bar{l}^{S^i*}\right\|_2},$$

where $ d_{norm}(\bar{l}^{S^i}_{d}, \bar{l}^{S^i*}) $ is a normalized Euclidean distance between the typical load profile of $ S^i$ at date $d$ and the actual typical load profile of device $ S^i$, $ (\bar{l}^{S^i})_{d} $ is a typical load profile of device $ S^i$ at date $d$, and $ \bar{l}^{S^i*} $ is a typical load profile of device $ S^i$ using all data.


Using the normalized Euclidean distance facilitates examining how fast our load profiles converge to their best estimator and solve the cold start problem. The optimal value for the normalized Euclidean distance is zero. To enable an easy comparison and analysis of the cold start problem, we need to summarize the cold start problem in one measure. We calculate the number of days each individual agent needs to reach a certain level of stability in its performance, reflecting the recommender system's cold start problem. According to our definition of stability, an agent's performance is stable if the performance satisfies a specified stability condition. For each day in the data set, we check if the stability condition is satisfied for the period starting from the day currently evaluated up to the last day. Hence, the time to solve the cold start problem is defined as the first day satisfying  the stability condition. 

For the Availability Agent and the Usage Agent, the AUC score must be within a specified tolerance interval around the best AUC score to satisfy the stability condition. The Load Agent's normalized Euclidean distance must be below a specified tolerance. The cold start problem of the complete framework is solved if the Availability Agent, the Usage Agent, and the Load Agent solve the cold start problem. 

\subsubsection{Evaluating the Performance of the Recommendation System}

The ultimate goal of our empirical analysis is to quantify how much a user can save by using our recommender system. To receive the potential energy cost saving of a recommendation, we create recommendations, determine their acceptability, and calculate the energy costs for running devices with and without our recommendations. We define a recommendation as acceptable if both the user availability and the device usage targets for the respective hour and day are positive. We calculate the potential cost savings based on the day-ahead energy prices provided by the Price Agent and the true device's energy consumption. Since we compute the cost savings for each provided recommendation, we aggregate our evaluation results to summarize the performance of our recommender system in a few metrics at the household level: the number of recommendations provided, the rate of acceptable recommendations, total savings, and relative savings.

\section{Empirical Results}
\label{EmpiricalResults}

\subsection{Data}

We use the REFIT Electrical Load Measurements data \citep{Murray2017} to analyze our recommender system. The data contains the energy consumption of nine different devices used in 20 households in the United Kingdom from 2013 to 2015. To validate our findings, we perform our evaluation steps on households 1 to 10. For the day-ahead prices provided by the Price Agent, we access the online database for industry day-ahead prices for the United Kingdom \citep{ENTSOE}. Even though these prices are not the actual prices that the households pay for their energy consumption, these prices indicate a realistic proxy for possible variable prices at the household level. Higher industry prices reflect higher demand in the market and result in higher energy prices for households with variable tariffs. Even without variable household prices, optimizing according to day-ahead wholesale electricity prices should help to match demand and supply better and consume more efficiently.


\subsection{Output of the Recommendation System}

We fit the  utility-based context-aware multi-agent recommendation system to the REFIT data. The tutorial in the Jupyter Notebook (link to be added after the review process) shows a detailed walkthrough on how to apply the proposed system.
The recommendation system generates a recommendation for each shiftable device within a household. The required input includes: the recommendation day, the availability threshold, and the device usage threshold.
Hours that have a smaller probability of user availability and the devices that have a smaller probability of usage are not considered for the recommendation. Finally, the system recommends the best launching hour on the recommendation date if both the activity and device usage flags are zero for a given device. 

We show an exemplary output of the recommendation system in Table \ref{tab:rs_output}. For the recommendation day (i.e., 15/02/2015) and three shiftable devices (i.e., tumble dryer, washing machine and dishwasher) the system calculates that the 8:00 a.m. is the cheapest hour to start operation. However, for the tumble dryer and the dishwasher the predicted probability of the device usage on this day is less than a threshold (i.e., device usage flag equals to one). Therefore, the user receives the recommendation for the washing machine only.

\vspace{0.5cm}
\begin{table}[h!]
    \centering
    \begin{tabular}{llllll}
        \toprule
        Recommendation  &	Device	& Best &	\multicolumn{1}{l}{Activity flag}  &	\multicolumn{1}{l}{Device usage flag} 	& \multicolumn{1}{l}{Final}  \\
         date &		&  hour & \multicolumn{1}{l}{(no recommendation)} &	 \multicolumn{1}{l}{(no recommendation)}	&  \multicolumn{1}{l}{recommendation} \\
        \midrule
        15/02/2015 &	Tumble Dryer  & 	8 &	0 &	1 &	no \\
        15/02/2015 &	Washing Machine &	8 &	0 &	0 &	8 \\
        15/02/2015 &	Dishwasher	   &    8 &	0 &	1 &	no \\
        \bottomrule
    \end{tabular}
    \vspace{0.5cm}
    \caption{\label{tab:rs_output} Exemplary output of the recommendation system.}
\end{table}

\subsection{Performance Evaluation}

This section aims to evaluate the performance of the applied recommendation system as described in Section \ref{PerformanceTheory}.
First, we evaluate the performance of individual agents.
Next, we identify the possible impact of the cold start problem on the agents through analyzing the system's performance over time.
Finally, we identify the optimal values for the hyperparameters and calculate the potential cost savings from the suggested recommendation system.

\subsubsection{Performance of Individual Agents}

To evaluate the performance of individual agents, we calculate the AUC scores for the Availability Agent and the Usage Agent and the MSE score for the Load Agent.
We evaluate the Availability Agent's performance on a household level, while the Usage Agent's and Load Agent's performance for each of three shiftable devices in each household. Table \ref{tab:individual-agent-scores} shows the results of the performance evaluation. Table \ref{tab:device-index-legend} in the Appendix describes shiftable devices for each household.

\vspace{5mm}
\begin{table}[h!]
    \centering
    \begin{tabular}{llrrrrrr}
        \toprule
        {} & Availability AUC & \multicolumn{3}{l}{Usage AUC} & \multicolumn{3}{l}{Load MSE} \\
         Devices &              &           &       &       &           &          &          \\
        Households &                 &         \multicolumn{1}{l}{0} &     \multicolumn{1}{l}{1} &     \multicolumn{1}{l}{2} &         \multicolumn{1}{l}{0} &        \multicolumn{1}{l}{1} &        \multicolumn{1}{l}{2} \\
        \midrule
        1         &         0.72 &      0.52 &  0.45 &  0.49 &   1224.45 &   511.49 &  3021.88 \\
        2         &         0.77 &      0.60 &   0.80 &     - &    320.81 &  14578.8 &        - \\
        3         &         0.80 &      0.65 &  0.65 &  0.65 &  24907.50 &   890.33 &   314.84 \\
        4         &         0.82 &      0.67 &  0.52 &     - &      0.80 &  1793.77 &        - \\
        5         &         0.76 &      0.68 &     - &     - &  37982.21 &        - &        - \\
        6         &         0.77 &      0.53 &  0.66 &     - &    424.78 &  6032.65 &        - \\
        7         &         0.83 &      0.79 &   0.80 &  0.81 &  43673.35 &   336.83 &  6845.26 \\
        8         &         0.62 &      0.77 &     - &     - &      0.35 &        - &        - \\
        9         &         0.67 &      0.60 &  0.57 &  0.78 &  30711.30 &    14.36 &  5480.13 \\
        10        &         0.77 &      0.74 &     - &     - &    647.57 &        - &        - \\
        \bottomrule
    \end{tabular}
    \vspace{0.5cm}
    \caption{\label{tab:individual-agent-scores}Performance of the individual agents.}
\end{table}


We do not detect any patterns in the data  explaining the variation of the Availability and Usage Agents' performance. Differences in users' behavior might drive these variations.  For instance, to which degree the user is active during the day and across different days, how frequently specific devices are used, or how often the user is absent, e.g., during vacations. 
The difference in the performance between the Availability Agent and the Usage Agent could also be due to different prediction intervals. While the Availability Agent predicts the availability for each hour of the day, the Usage Agent predicts the device usage daily. It might be easier for the model to detect hourly patterns that occur over a day than daily patterns that occur over a week or a month. 
The Load Agent’s performance variance is due to the way of calculation of the typical load profile and the user's usage behavior. Using the concept of a typical load profile is appropriate for the approximation of the energy consumption of devices with the low duration and/or consumption variability (e.g., a dishwasher used on similar programs). In contrast, if a device has different programs or the duration time is chosen by the user, the typical load profile yields a large MSE score. For instance, the very large MSE scores for the device 0 at the households 3, 5, 7 and 9 belong to the tumble dryers. This can be explained by the variability of the programs, but also by the fact that the laundry characteristics influence the operation's duration of the tumble dryers (e.g., type of material, material thickness and dryness level). The smallest MSE levels in Table \ref{tab:individual-agent-scores} correspond to the washing machines.

\subsubsection{Cold Start Problem Evaluation}

\begin{figure}[h!]
	\centering
		\includegraphics[width=1\textwidth]{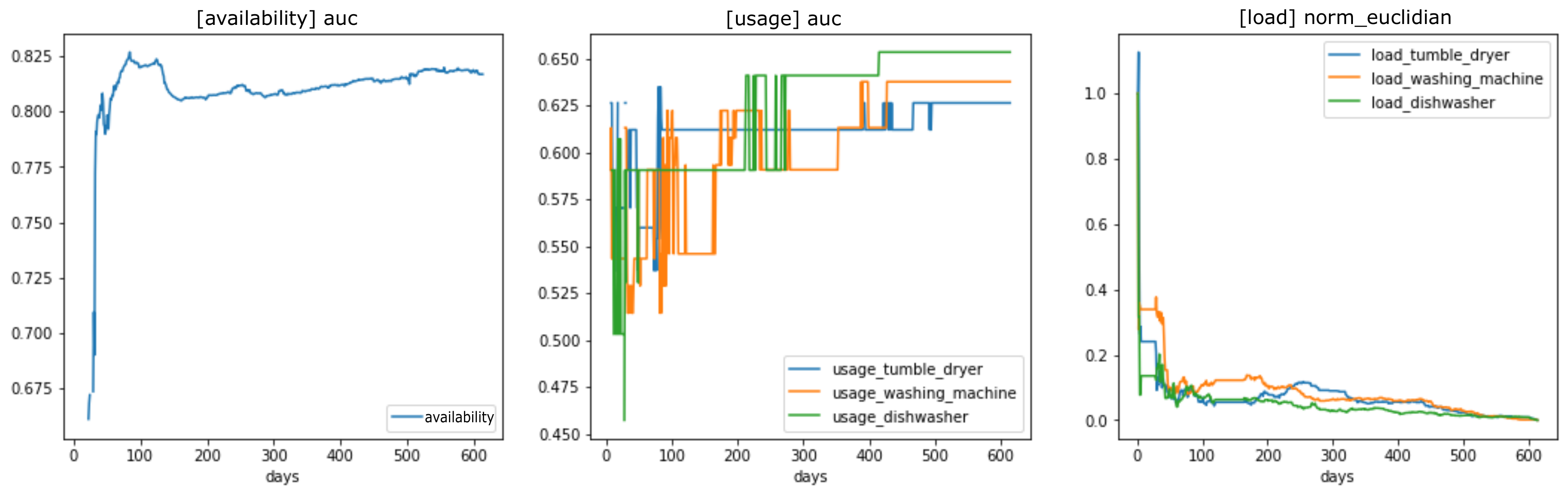}
	\caption{Performance of the Availability Agent, the Usage Agent and the Load Agent over time for the household 3 for the cold start problem evaluation.}
	\label{fig:cold-start_1}
\end{figure}
\begin{figure}
	\centering
		\includegraphics[width=1\textwidth]{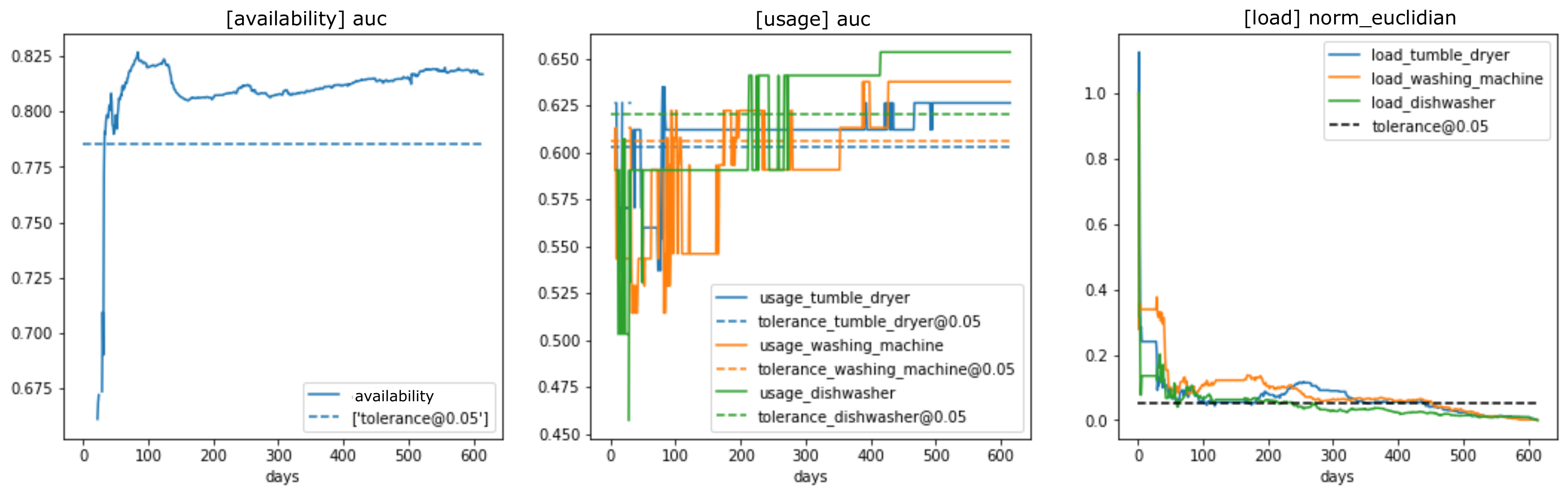}
		\includegraphics[width=1\textwidth]{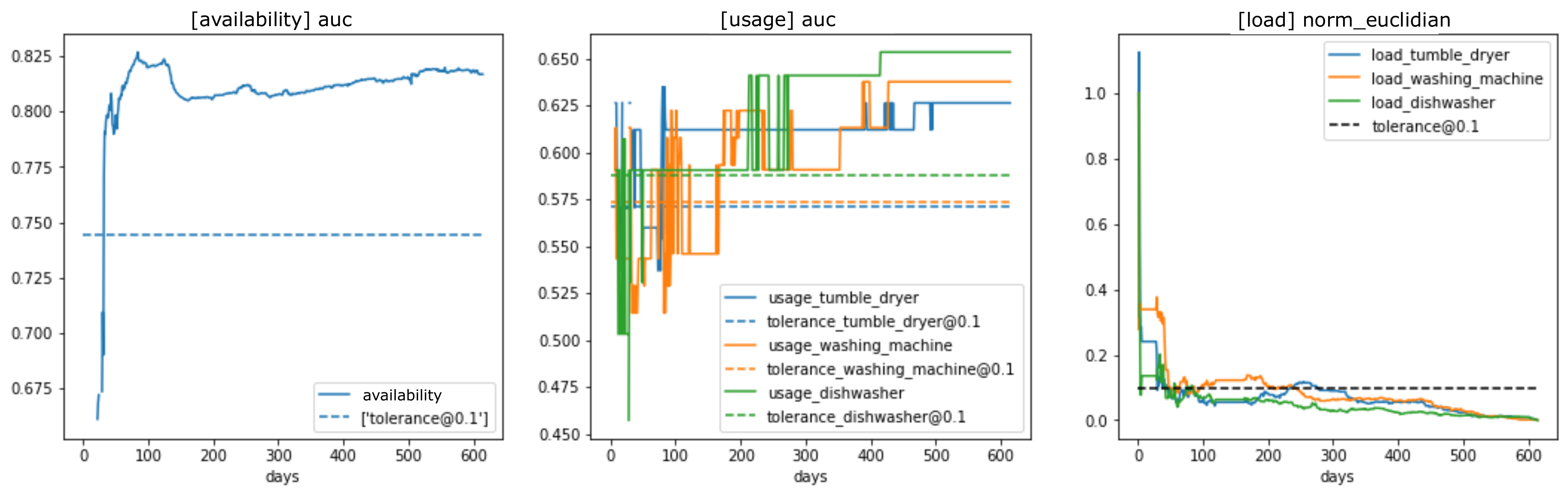}
		\includegraphics[width=1\textwidth]{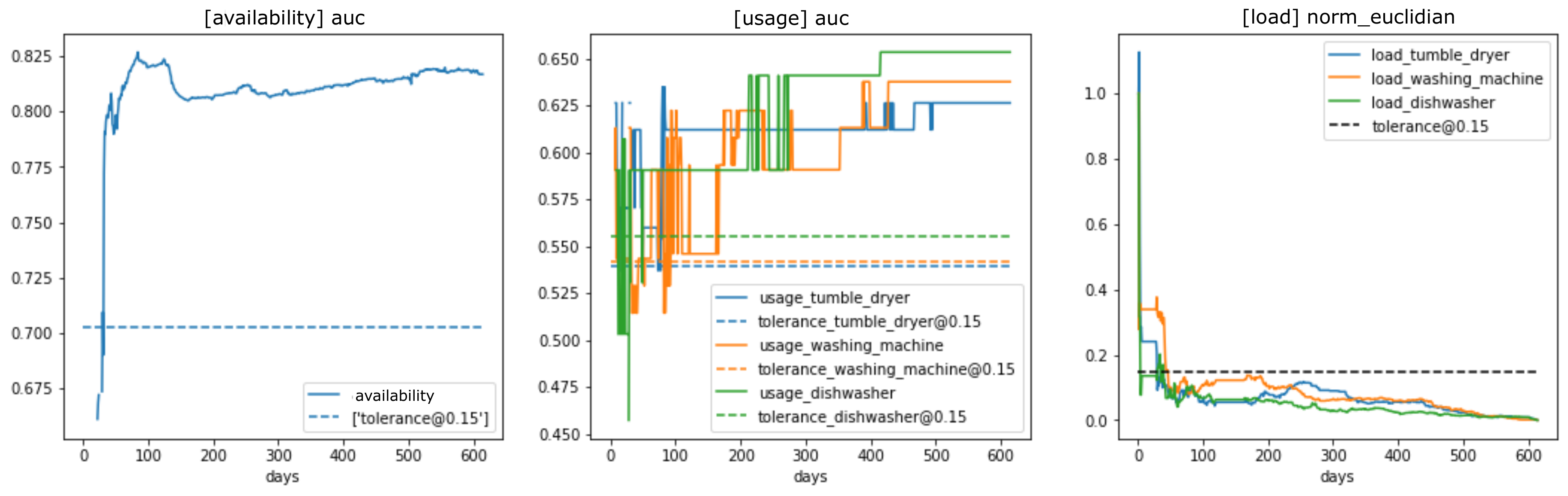}
	\caption{Different tolerance values applied to household 3.}
	\label{fig:cold-start_2}
\end{figure}

Further on we continue with evaluating each agent for varying lengths of data. For instance, Figure \ref{fig:cold-start_1} shows the performance of the Availability Agent, the Usage Agent and the Load Agent over time for the household 3.
We test different tolerance values (see Figure \ref{fig:cold-start_2}) to compute the duration and solve the cold start problem. We select the most appropriate tolerance level (0.15) using a visual inspection.
Table \ref{tab:cold-start-days} presents a solution of the cold start problem for the ten households.
The Availability Agent seems not to suffer from the extreme cold start problem and the cold start days are quite constant across different households. 
In contrast to the Availability Agent, the cold start scores for the Usage Agent vary heavily across devices and households.
This suggests that predicting daily  patterns might be harder than predicting hourly ones.
Stable performance and hence a low number of cold start days of the Usage Agent would indicate that users operate their devices according to some patterns or habits on a daily basis. The varying cold start scores for the Usage Agent suggest that users do not use the devices according to such patterns or often change their device usage behavior.
The cold start score for the Load Agent implies how much time the device's load profile needs to converge to its best estimator. 
For instance, the value of the cold start days for the tumble dryer (device 0) in the household 3 equals to 36 (see \ref{tab:cold-start-days}). This means that the length of the training data for the household 3 should be at least 36 days to provide appropriate recommendations for the tumble dryer.
Similar to the individual agents' scores, a weak Load Agent performance might be due to the user changing the usage behavior or due to the device consuming energy differently in varying contexts.
The last column in Table \ref{tab:cold-start-days} shows the solution of the cold start problem. For instance, for the household 3 the minimum period needed for the three agents to solve the cold start problem is 92 days. Thus, the complete framework is solved in 92 days.
To sum up, the solutions of the cold start problem show high variability over different households. Again, we see the differences in energy consumption behavior across households as a main driver of this variability.


\begin{table}
    \centering
    \begin{tabular}{llrrrrrrr}
        \toprule
        {} & \multicolumn{1}{l}{Availability} & \multicolumn{3}{l}{Usage} & \multicolumn{3}{l}{Load} & Solution \\
        Devices &          &     \multicolumn{1}{l}{0} &    \multicolumn{1}{l}{1} &    \multicolumn{1}{l}{2} &    \multicolumn{1}{l}{0} &    \multicolumn{1}{l}{1} &    \multicolumn{1}{l}{2} &          \\
        Households &          &       &      &      &      &      &      &           \\
        \midrule
        1         &       19 &   275 &  639 &  639 &  309 &   16 &   21 &       639 \\
        2         &       10 &   618 &   24 &    - &  230 &    5 &    - &       618 \\
        3         &       32 &    78 &   92 &   50 &   36 &   47 &   40 &        92 \\
        4         &       11 &   205 &   49 &    - &  311 &  177 &    - &       311 \\
        5         &       11 &    27 &    - &    - &  429 &    - &    - &       429 \\
        6         &       11 &    44 &  573 &    - &   54 &  288 &    - &       573 \\
        7         &       11 &   159 &   78 &  198 &  210 &  129 &  229 &       229 \\
        8         &       21 &    81 &    - &    - &  132 &    - &    - &       132 \\
        9         &       13 &   568 &  110 &   11 &  256 &  141 &   19 &       568 \\
        10        &       31 &    95 &    - &    - &   23 &    - &    - &        95 \\
        \bottomrule
    \end{tabular}
    \vspace{0.5cm}
    \caption{\label{tab:cold-start-days}Cold start days at the tolerance value of 0.15 for households 1 to 10.}
\end{table}

\subsubsection{Performance of the Recommendation System}

For creating recommendations, we need to specify two hyperparameters: availability threshold and usage threshold. These are initiated by the system and can be adjusted by the user. To initialize them optimally, we analyze the sensitivity of recommendation timing and our performance measures to changes in these hyperparameters. Additionally, we perform a grid search over the candidate thresholds to identify the optimal values. 

\begin{figure}[h!]
	\centering
		\includegraphics[width=1\textwidth]{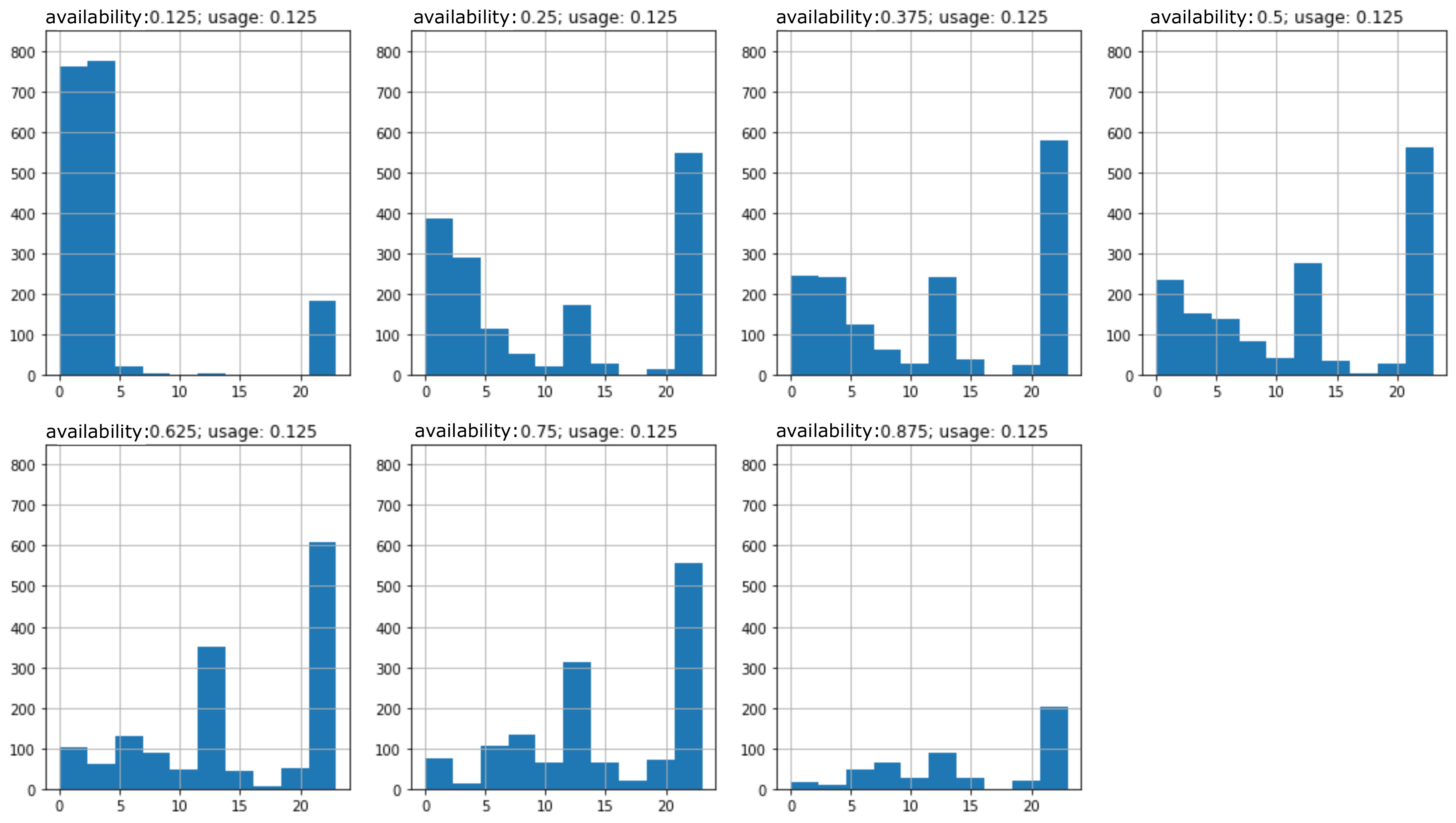}
	\caption{Changes in recommended starting hours for the household 3 regarding different availability thresholds and a constant usage threshold.}
	\label{fig:timing-1}
\end{figure}

\begin{figure}[h!]
	\centering
		\includegraphics[width=1\textwidth]{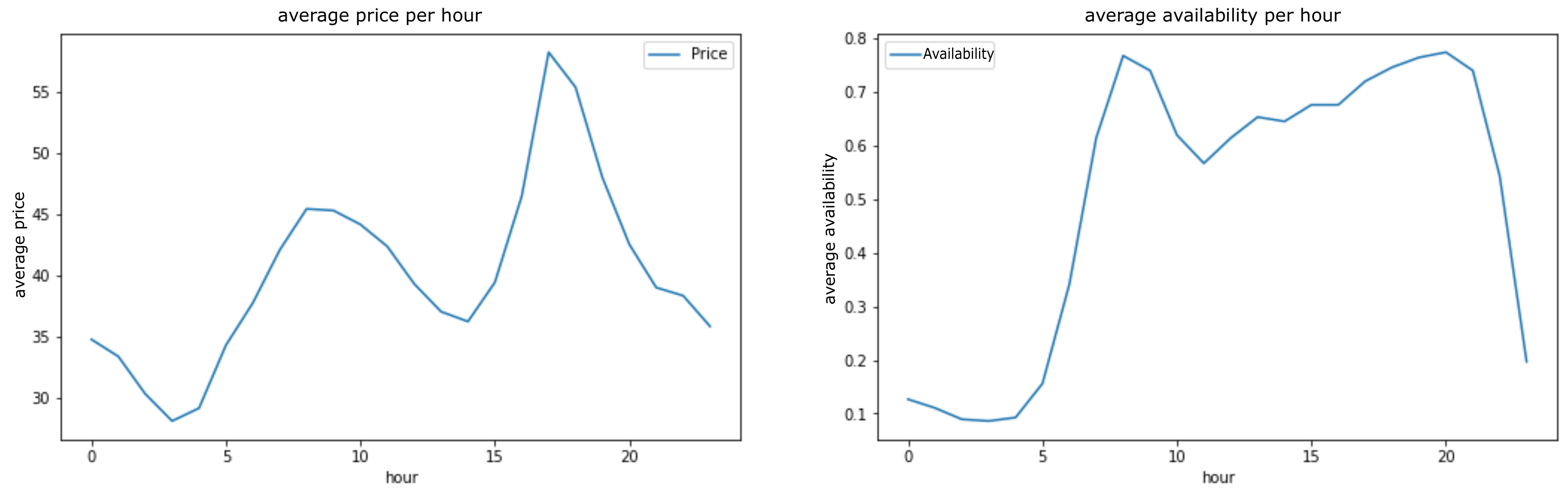}
	\caption{Average day-ahead prices per hour and average availability per hour for the household 3.}
	\label{fig:timing-2}
\end{figure}

We analyze the recommendation timing to determine how changes in the recommendation parameters shift recommendations to different hours of the day. Since the system provides the device usage predictions daily, the usage threshold does not influence the hour of the recommendation. Therefore, we focus on the availability threshold and set the usage threshold to a constant value for the recommendation timing analysis. When comparing the recommendation distribution over the 24 hours for various availability thresholds (see Figure \ref{fig:timing-1}) with the average energy price per day and the average user availability per hour (see Figure \ref{fig:timing-2}), we observe the expected behavior. For lower availability thresholds, the system focuses more on prices rather than on the user's availability. Hence, setting the low availability threshold leads to generating most recommendations for the hours with low energy costs. In contrast, setting higher availability threshold results in recommended starting hours that are more in line with the user's availability. It should be noted that setting the availability threshold too high could even lead to losses. If the recommender system focuses too much on the user's availability, it can reduce the potential recommendation hours only to a few hours per day in which the user is very likely to be active. The prices for these hours might be higher than at the otherwise recommended hour.

To find the optimal recommendation hyperparameters, we perform a grid search over the candidate thresholds. We determine the optimal recommendation hyperparameters during the grid search as a parameter combination that maximizes the total energy cost savings. While the calculated total savings value might not be useful for the household, comparing various scales of total savings for different hyperparameter combinations enables finding a parameter combination that allows the highest total savings at the household level. 

\begin{figure}
	\centering
		\includegraphics[width=1\textwidth]{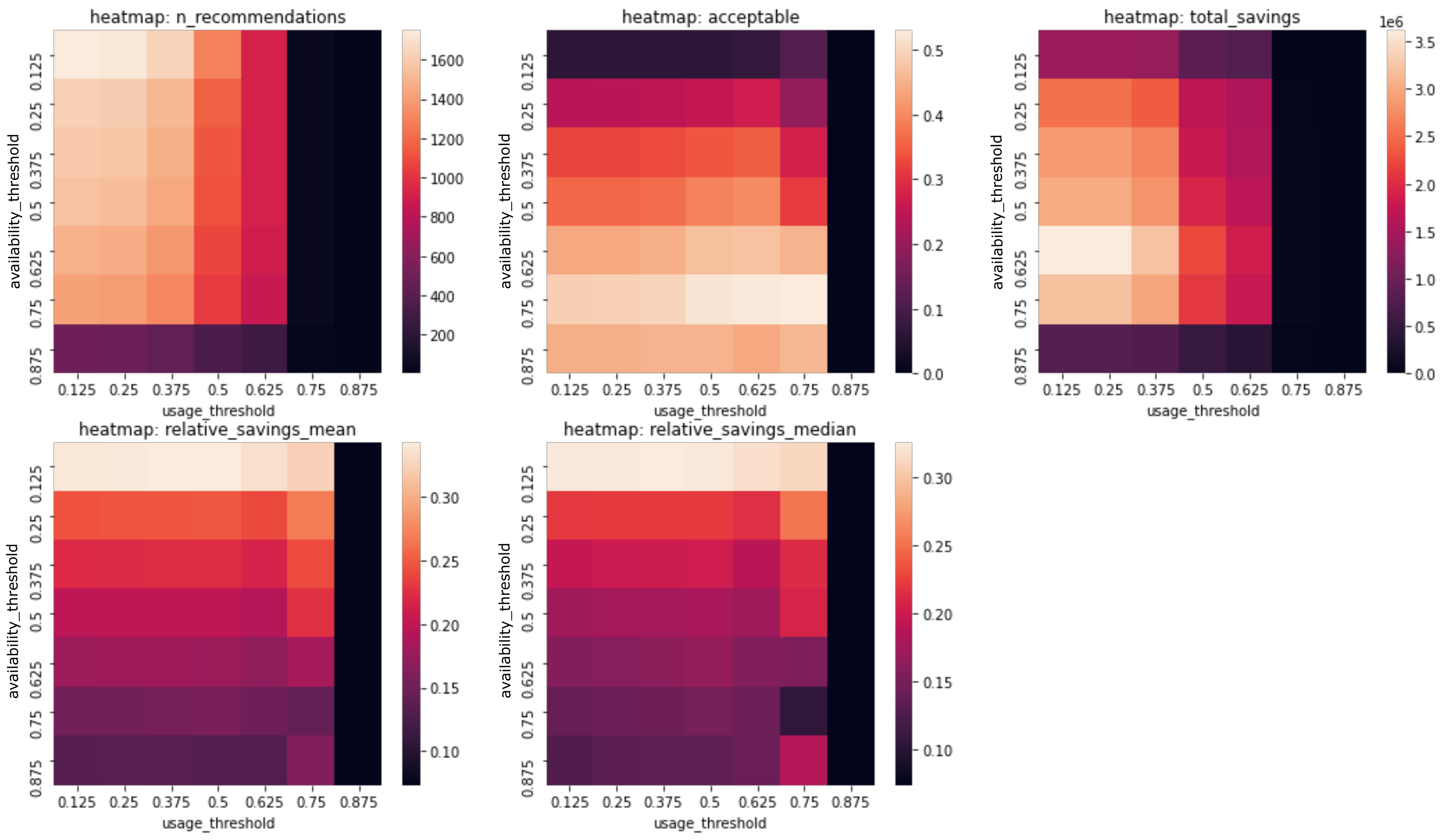}
	\caption{Sensitivity of performance metrics to changes in the hyperparameters for the household 3.}
	\label{fig:sensitivity-1}
\end{figure}

To evaluate the potential energy cost savings, we summarize the performance of the recommendation system with the following metrics at the household level: the number of recommendations provided, the rate of acceptable recommendations, total savings, and relative savings. 
The grid search results facilitate the analysis of sensitivity of the performance metrics to changes in the hyperparameters (see Figure \ref{fig:sensitivity-1}). 
The results from the sensitivity analysis also coincide with the expected behavior. The \textbf{number of recommendations} is the highest for the low threshold values while setting the thresholds too high leads to few or no recommendations. This behavior holds for both, the availability threshold and the usage threshold. However, the sensitivity is stronger for the usage threshold. This might be because the availability threshold applies to multiple potential recommendation hours, while the usage threshold only to the recommendation day. Hence, it is more likely that at least one availability probability for a potential starting hour exceeds the availability threshold compared to the single usage probability exceeding the usage threshold. 
In contrast to the number of recommendations, the \textbf{rate of acceptable recommendations} increases with increasing threshold values. However, setting the thresholds to high values results in reducing the rate of acceptable recommendations or even preventing acceptable recommendations.
For the \textbf{relative savings}, the main driver is the availability threshold. The availability threshold determines the trade-off between focusing on the price structure or the user's availability. A low availability threshold allows recommendations with a high cost-saving potential, as these recommendations are provided mainly for hours when the price is the lowest. Thus, if the recommendations are acceptable, the relative savings are high. Only a high value of the usage threshold influences the relative savings. In this case, the system generates no recommendations, and relative savings drop to zero. 
The sensitivity of the \textbf{total savings} takes all the effects mentioned above into account. The availability threshold determines the trade-off between acceptable recommendations and the total savings. The usage threshold should be set relatively low to maximize the total savings in absolute terms and not to miss cost-saving opportunities. In such case, the system produces a high number of recommendations.

Table \ref{tab:performance-metrics} shows the performance measures across households. The optimal threshold values in terms of cost savings for different households represent various trade-offs mentioned in the sensitivity analysis above. 
Here we observe two main trends. First, the  value of the availability threshold tends to be higher. Thus, the recommender system focuses more on the users' availability rather than on low prices to maximize the users' total savings. Second, for the usage threshold, it seems to be optimal to choose lower threshold values. This is because the chance of missing out on a potential savings opportunity is low with a low usage threshold. However, the current parameter tuning strategy does not account for any disutility of receiving too many recommendations or not acceptable recommendations. When accounting for this disutility, we assume the optimal usage threshold would increase, the number of recommendations would decrease, and the rate of acceptable recommendations would increase.

\begin{table}
    \centering
    \begin{tabular}{rrrrrrr}
        \toprule
        Households & Availability th. & Usage th. &  Acceptable &  \# Recom. &  Rel. savings &  Total savings \\
        \midrule
        1  &              0.375 &           0.125 &        0.15 &              959 &                   0.10 &      254,308 \\
        2  &              0.500 &           0.125 &        0.38 &              875 &                   0.18 &     2,431,697 \\
        3  &              0.625 &           0.125 &        0.44 &             1484 &                   0.18 &     3,613,360 \\
        4  &              0.750 &           0.125 &        0.27 &              822 &                   0.20 &      589,320 \\
        5  &              0.750 &           0.250 &        0.52 &              269 &                   0.15 &     2,258,746 \\
        6  &              0.375 &           0.125 &        0.14 &              713 &                   0.18 &      502,190 \\
        7  &              0.625 &           0.125 &        0.57 &             1404 &                   0.18 &     6,676,910 \\
        8  &              0.125 &           0.125 &        0.32 &              519 &                   0.03 &        1,071 \\
        9  &              0.375 &           0.125 &        0.21 &             1364 &                   0.21 &     2,611,549 \\
        10 &              0.500 &           0.125 &        0.31 &              436 &                   0.20 &      365,033 \\
\bottomrule
\end{tabular}
    \vspace{0.5cm}
    \caption{\label{tab:performance-metrics}Recommender system performance for the households 1 to 10.}
\end{table}

Concerning the rate of acceptable recommendations, the optimal thresholds from the grid search support our findings from the sensitivity analysis. The threshold values and the acceptance rates seem to be positively correlated. 
However, we do not receive the same results for the number of recommendations detected during the sensitivity analysis. This might be mainly since  the households' number of shiftable devices and the individual agents' performance differ across the households.
While analyzing the potential energy cost savings, we observe the expected behavior for the total savings in absolute terms. These vary heavily across different households, as many factors play a role in absolute savings. These factors might include the number of shiftable devices, the performance of individual agents, the energy consumption of individual devices as well as how efficient the user's behavior is without the recommender system. It is remarkable that despite various factors that play a role in determining the savings, most of households could benefit 18 and more percent of relative savings. 
An interesting case is a household 8. The optimal availability threshold is very low compared to the rest. This means that the recommendations are more focused on lower prices than the users' availability. However, the acceptance rate does not react as strongly to the small threshold values as for other households. According to our sensitivity analysis, combining these two effects should lead to higher relative and total savings. Nevertheless, for the household 8, this is not the case. This could indicate that the household already uses its shiftable devices quite efficiently.

\section{Discussion}
\label{Discussion}


This section discusses the suggested approach by pointing out its strengths, weaknesses, and potential improvements. The feasibility of our technical implementation is supported by the previous work of \cite{JimenezBravo2019}. They provide an architecture of communicating agents that solves various sub-tasks contributing to a final recommendation. Both approaches use similar context data, have the same hardware requirements, and possess a similar recommendation logic. 
We see our works as complementary since \cite{JimenezBravo2019} focus on the working architecture, whereas we focus more on the recommendation algorithm. In addition, the design of our recommender system allows it to be easily integrated into the architecture of \cite{JimenezBravo2019}. Nevertheless, our approach improves in several dimensions.

First, \cite{JimenezBravo2019} and our work account for device usage habits to provide recommendations when they are the most relevant. For this purpose, \cite{JimenezBravo2019} provide a single recommendation per device per week on the weekday of most frequent historical device use. In contrast, our approach is statistically grounded and uses a probability of the device usage on a given day to decide whether to recommend  or not. As a result, our recommender system flexibly adapts to actual usage patterns or changing usage behavior. Moreover, it can provide multiple recommendations per week for frequently used devices and few (or none) for infrequently used ones.

Second, our algorithm allows calculating costs with higher precision. Indeed \cite{JimenezBravo2019} only consider electricity prices at the device starting hour, ignoring device load profiles. Thus, if a user operates a device for more than one hour, this approach is problematic since high prices might follow immediately after low prices. For example, if a washing machine runs for three hours, their system could recommend starting the device at an hour with a low electricity price, without considering that the two following usage hours could be much more expensive. Our system takes into account devices' load profiles to compute an overall estimation of the usage costs. 

Lastly, \cite{JimenezBravo2019} aim at making recommendations at hours when the user is active. However, their paper does not specify the exact solution for this approach. In contrast, we provide a fully functional and statistically grounded solution for predicting user availability. 

In contrast to the existing approaches 
\citep{Paunescu2014, Han2009}, our paper deals with a recommender system that is based on the end consumers' willingness and own action to shift the demand. In that context, end-consumers might choose different goals over which to optimize their consumption. An obvious option is to save costs, as mentioned in \citet{JimenezBravo2019}. Thus, consumers can shift the use of electrical appliances to times when the price of electricity is lowest and when they are at home. To summarize, it is up to them to decide whether to accept or reject the received recommendations.

We see the main strength of the recommendation system in its pragmatic simplicity. For the deployment, the system needs ordinary data and a minimal user input. In terms of hardware, our system requires smart plugs and an interface for the recommendations (e.g., on a smartphone). The architecture provides a model that computes quickly, can be implemented locally and does not require sharing highly sensitive data such as device usage consumption. This makes the approach realistically implementable in residential buildings. 
Practically, it could be integrated into a smart home system or implemented as a standalone smartphone application that provides daily recommendations for device usage starting times. In that way, households could save costs and better achieve their sustainability goals.
Since the system optimizes the electricity usage, the households should also contribute to flattening demand peaks on the electricity market. At the same time, it accounts for important user preference aspects such as device usage habits, costs and user availability.


\subsection{Implications}

We see the following relevant societal impacts and potential side-effects. Firstly, our recommendation system strongly encourages changes in energy consumption behavior. Although the system accounts for the user's daily device usage habits (for making recommendations on days when the usage probability is high), it does not consider the hours of typical device usage. Instead, it encourages a user to shift the device usage to hours of efficient consumption within hours of likely user availability. As a result, the generated recommendations facilitate more radical changes towards energy-efficient consumption. The possible drawback lies in more likely rejections of recommendations. 

Secondly, we consider the lack of data as a significant challenge in the evolution of recommender systems for energy efficiency and load shifting. The availability of simple and ready-to-use recommender systems like ours could encourage households to install smart plugs. More smart plugs would facilitate collecting data on energy consumption behavior that is currently very scarce. The data collected by a simpler model could enable the development of more complex and advanced approaches.

Finally, using recommender systems for load shifting can prove to be a valuable instrument for facilitating the transition to a renewable energy supply. Our current implementation aims at total energy cost savings. However, the target can be easily switched to maximize the proportion of renewable energy in the devices' electricity consumption. For instance, the system would recommend using appliances during hours with a higher green energy supply in the electricity grid (provided the predicted user availability and the probability of the device usage are high enough).

\subsection{Limitations}

The simplicity of the approach does not come without a cost. Using only local historical device consumption data enables simple prediction models that might lack accuracy when facing complex patterns. For instance, recommendations might not account well varying device usage patterns. In the current implementation, our system generates only one recommendation per device per day. Therefore, it cannot adapt well to devices that are used frequently within a day. Similarly, since the usage cost computation relies on a typical load profile, the recommender system cannot adapt well to devices with varying usage duration or electricity load profiles.

Another issue might reside in the way our approach determines user availability. We assume that if the energy consumption exceeds a certain threshold, the user operates the device. In other words, the user is available to implement recommendations. In addition, the devices' energy consumption can be accompanied by random noise that we can not detect during data cleaning. This can lead to a weaker performance of the model. Furthermore, the user specifies in advance appliances indicating the availability. Thus, we assume that the user is available if these appliances are turned on. However, it might be possible that our assumption is violated. A device might have some automatic reoccurring patterns (e.g., reboot once a day or week), which we can not isolate and clean in the data.  Given the labeled data, one can easily predict these reoccurring patterns. Taking these predictions as additional input parameters could boost the performance of the Availability Agent or the Usage Agent. Another approach to deal with this issue could be using motion sensor data to track the user availability (however, at the price of complexity).

\subsection{Future Work}

Although simplicity is a decisive factor in the design of our approach, there are many directions for further improvement. For example, users need to indicate which devices are shiftable and which devices show user availability. These tasks could be automated. For instance, it might be possible to train an algorithm to recognize devices based on their load profiles and determine whether such a device is suitable for recommendations or availability tracking. Next, the user can currently adjust the probability thresholds manually. However, this task could also be automated (for example, by adjusting thresholds using automated recommendation feedback, i.e., tracking when recommendations are most often implemented). 
Finally, we use rather simple models to predict or approximate the users' behavior or preferences within our agents. The agents' performance could be increased by using more sophisticated models for the machine learning tasks. 
Fortunately, due to the multi-agent architecture, such changes could be easily integrated into our general recommendation framework by adapting individual agents without impacting the general logic of recommendations. This flexibility also allows to easily change the target used for optimizing recommendations (for instance, environmental costs such as carbon emissions per production unit instead of total energy cost savings).



\section{Conclusions}
\label{Conclusions}

This paper introduces a utility-based context-aware recommendation system for energy efficiency via load shifting in residential buildings and proposes a framework for its
performance evaluation. 
We provide a solution with low requirements for data, user input, hardware, and computational resources. In addition, the system does not require sharing highly sensitive data such as device usage electricity consumption. The recommendation algorithm can be integrated into a smart home system or implemented as a standalone smartphone application that provides daily recommendations for device usage starting times. In this way, households could save costs, optimize individual electricity usage, contribute to the stability of the power grid, and, as a result, reduce the CO\textsubscript{2} emissions.

\bibliographystyle{unsrtnat}

\bibliography{references} 

\clearpage

\appendix

 \newpage%
 \renewcommand{\thesection}{\Alph{section}}
 \section{Appendix}

\vspace{0.5cm}
\begin{table}[h!]
    \centering
    \begin{tabular}{llll}
        \toprule
                  &                     &                 devices &            \\
        household &                    0  &                   1  &          2  \\
        \midrule
        1         &         Tumble Dryer &      Washing Machine &  Dishwasher \\
        2         &      Washing Machine &           Dishwasher &           - \\
        3         &         Tumble Dryer &      Washing Machine &  Dishwasher \\
        4         &  Washing Machine (1) &  Washing Machine (2) &           - \\
        5         &         Tumble Dryer &                    - &           - \\
        6         &      Washing Machine &           Dishwasher &           - \\
        7         &         Tumble Dryer &      Washing Machine &  Dishwasher \\
        8         &      Washing Machine &                    - &           - \\
        9         &         Washer Dryer &      Washing Machine &  Dishwasher \\
        10        &      Washing Machine &                    - &           - \\
        \bottomrule
    \end{tabular}
    \vspace{0.5cm}
    \caption{\label{tab:device-index-legend}Legend for mapping the shiftable devices to an integer index.}
\end{table}
\vspace{5mm}

  

\end{document}


\maketitle

\begin{abstract}
	TBA
\end{abstract}

\keywords{Recommendation System \and Energy Efficiency \and Load Shifting \and Energy Consumption Behavior \and Smart Home}

\newpage

\section*{Appendix}

\begin{table}[h!]
    \caption{\label{tab:device-index-legend}Legend for mapping the shiftable devices to an integer index.} 

\vskip 0.15in
\centering
    \begin{tabular}{llll}
        \toprule
                  &                     &                 devices &            \\
        household &                    0  &                   1  &          2  \\
        \midrule
        1         &         Tumble Dryer &      Washing Machine &  Dishwasher \\
        2         &      Washing Machine &           Dishwasher &           - \\
        3         &         Tumble Dryer &      Washing Machine &  Dishwasher \\
        4         &  Washing Machine (1) &  Washing Machine (2) &           - \\
        5         &         Tumble Dryer &                    - &           - \\
        6         &      Washing Machine &           Dishwasher &           - \\
        7         &         Tumble Dryer &      Washing Machine &  Dishwasher \\
        8         &      Washing Machine &                    - &           - \\
        9         &         Washer Dryer &      Washing Machine &  Dishwasher \\
        10        &      Washing Machine &                    - &           - \\
        \bottomrule
    \end{tabular}

\end{table}
\vspace{5mm}

\begin{table}[h!]
    \caption{\label{tab:individual-agent-scores}Performance of the individual agents.}    
\vskip 0.15in
    \centering
    \begin{tabular}{lrrllrll}
        \toprule
        {} & availability\_auc & \multicolumn{3}{l}{usage\_auc} & \multicolumn{3}{l}{load\_mse} \\
         {} &              &           &       &   devices    &           &          &          \\
        household &                 &         0 &     1 &     2 &         0 &        1 &        2 \\
        \midrule
        1         &         0.72 &      0.52 &  0.45 &  0.49 &   1224.45 &   511.49 &  3021.88 \\
        2         &         0.77 &      0.60 &   0.8 &     - &    320.81 &  14578.8 &        - \\
        3         &         0.80 &      0.65 &  0.65 &  0.65 &  24907.50 &   890.33 &   314.84 \\
        4         &         0.82 &      0.67 &  0.52 &     - &      0.80 &  1793.77 &        - \\
        5         &         0.76 &      0.68 &     - &     - &  37982.21 &        - &        - \\
        6         &         0.77 &      0.53 &  0.66 &     - &    424.78 &  6032.65 &        - \\
        7         &         0.83 &      0.79 &   0.8 &  0.81 &  43673.35 &   336.83 &  6845.26 \\
        8         &         0.62 &      0.77 &     - &     - &      0.35 &        - &        - \\
        9         &         0.67 &      0.60 &  0.57 &  0.78 &  30711.30 &    14.36 &  5480.13 \\
        10        &         0.77 &      0.74 &     - &     - &    647.57 &        - &        - \\
        \bottomrule
    \end{tabular}

\end{table}


\begin{figure}[h!]
\vskip 0.2in
	\centering
		\includegraphics[width=0.9\textwidth]{./Bilder/cold-start-no-tolerance_h3.pdf}
	\caption{Performance of the Availability Agent, Usage Agent and Load Agent over time for household 3 for the cold start problem evaluation.}
	\label{fig:cold-start_1}
\end{figure}

\begin{figure}
\vskip 0.2in
	\centering
		\includegraphics[width=0.9\textwidth]{./Bilder/cold-start-tolerance-5_h3.pdf}
		\includegraphics[width=0.9\textwidth]{./Bilder/cold-start-tolerance-10_h3.pdf}
		\includegraphics[width=0.9\textwidth]{./Bilder/cold-start-tolerance-15_h3.pdf}
	\caption{Different tolerance levels applied to household 3.}
	\label{fig:cold-start_2}
\end{figure}

\vspace{1cm}
\begin{table}
    \caption{\label{tab:cold-start-days}Cold start days at the tolerance value of 0.15 for households 1 to 10.}
    \vskip 0.15in
    \centering
    \begin{tabular}{lrrllrllr}
        \toprule
        {} & availability & \multicolumn{3}{l}{usage} & \multicolumn{3}{l}{load} & framework \\
        device &          &     0 &    1 &    2 &    0 &    1 &    2 &         - \\
        household &          &       &      &      &      &      &      &           \\
        \midrule
        1         &       19 &   275 &  639 &  639 &  309 &   16 &   21 &       639 \\
        2         &       10 &   618 &   24 &    - &  230 &    5 &    - &       618 \\
        3         &       32 &    78 &   92 &   50 &   36 &   47 &   40 &        92 \\
        4         &       11 &   205 &   49 &    - &  311 &  177 &    - &       311 \\
        5         &       11 &    27 &    - &    - &  429 &    - &    - &       429 \\
        6         &       11 &    44 &  573 &    - &   54 &  288 &    - &       573 \\
        7         &       11 &   159 &   78 &  198 &  210 &  129 &  229 &       229 \\
        8         &       21 &    81 &    - &    - &  132 &    - &    - &       132 \\
        9         &       13 &   568 &  110 &   11 &  256 &  141 &   19 &       568 \\
        10        &       31 &    95 &    - &    - &   23 &    - &    - &        95 \\
        \bottomrule
    \end{tabular}

\end{table}

\begin{figure}[h!]
\vskip 0.2in
	\centering
		\includegraphics[width=0.9\textwidth]{./Bilder/timing-1_h3.pdf}
	\caption{Changes in recommended starting hours for household 3 regarding different availability thresholds and a constant usage threshold.}
	\label{fig:timing-1}
\end{figure}

\vspace{5mm}

\begin{figure}[h!]
\vskip 0.2in
	\centering
		\includegraphics[width=0.9\textwidth]{./Bilder/timing-2_h3.pdf}
	\caption{Average day-ahead prices per hour and average availability per hour for household 3.}
	\label{fig:timing-2}
\end{figure}

\begin{figure}
\vskip 0.2in
	\centering
		\includegraphics[width=0.9\textwidth]{./Bilder/sensitivity_h3.png}
	\caption{Sensitivity of performance metrics to changes in hyperparameters for the household 3.}
	\label{fig:sensitivity-1}
\end{figure}

\begin{table}
    \caption{\label{tab:performance-metrics}Recommender system performance for households 1 to 10.}
\vskip 0.15in
    \centering
    \begin{tabular}{ccccrcr}
        \toprule
        households & availability th. & usage th. &  acceptable &  \# recom. &  rel. savings &  total savings \\
        \midrule
        1  &              0.375 &           0.125 &        0.15 &              959 &                   0.10 &      254,308 \\
        2  &              0.500 &           0.125 &        0.38 &              875 &                   0.18 &     2,431,697 \\
        3  &              0.625 &           0.125 &        0.44 &             1484 &                   0.18 &     3,613,360 \\
        4  &              0.750 &           0.125 &        0.27 &              822 &                   0.20 &      589,320 \\
        5  &              0.750 &           0.250 &        0.52 &              269 &                   0.15 &     2,258,746 \\
        6  &              0.375 &           0.125 &        0.14 &              713 &                   0.18 &      502,190 \\
        7  &              0.625 &           0.125 &        0.57 &             1404 &                   0.18 &     6,676,910 \\
        8  &              0.125 &           0.125 &        0.32 &              519 &                   0.03 &        1,071 \\
        9  &              0.375 &           0.125 &        0.21 &             1364 &                   0.21 &     2,611,549 \\
        10 &              0.500 &           0.125 &        0.31 &              436 &                   0.20 &      365,033 \\
\bottomrule
\end{tabular}
\end{table}

\begin{table}[h!]
    \caption{\label{tab:rs_output} Exemplary output of the recommendation system.}
\vskip 0.15in
    \centering
    \begin{tabular}{llllll}
        \toprule
        Recommendation  &	Device	& Best   &	\multicolumn{1}{l}{Availability flag}  &	\multicolumn{1}{l}{Device usage flag} 	& \multicolumn{1}{l}{Final}  \\
         date &		&  hour & \multicolumn{1}{l}{(no recommendation)} &	 \multicolumn{1}{l}{(no recommendation)}	&  \multicolumn{1}{l}{recommendation} \\
        \midrule
        15/02/2015 &	Tumble Dryer  & 	8 &	0 &	1 &	no \\
        15/02/2015 &	Washing Machine &	8 &	0 &	0 &	8 \\
        15/02/2015 &	Dishwasher	   &    8 &	0 &	1 &	no \\
        \bottomrule
    \end{tabular}
\end{table}
